\documentclass[10pt,conference]{IEEEtran}

\usepackage[font={footnotesize}]{caption} 

\usepackage{graphicx} 
\usepackage{url} 
\usepackage{paralist} 
\usepackage{booktabs} 

\usepackage{tikz}
\usepackage{textcomp}

\IEEEoverridecommandlockouts 
\usepackage{siunitx} 
\usepackage{hyperref} 

\newcommand\copyrighttext{%
  \footnotesize \textcopyright 2022 IEEE. Personal use of this material is permitted. Permission from IEEE must be obtained for all other uses, in any current or future media, including reprinting/republishing this material for advertising or promotional purposes, creating new collective works, for resale or redistribution to servers or lists, or reuse of any copyrighted component of this work in other works.
  }
\newcommand\copyrightnotice{%
\begin{tikzpicture}[remember picture,overlay]
\node[anchor=south,yshift=10pt] at (current page.south) {\fbox{\parbox{\dimexpr\textwidth-\fboxsep-\fboxrule\relax}{\copyrighttext}}};
\end{tikzpicture}%
}

\begin{document}

\title{Reproducible Cross-border\\  High Performance Computing
for Scientific Portals}

\author{
	\IEEEauthorblockN{Kessy Abarenkov\IEEEauthorrefmark{1}, Anne Fouilloux\IEEEauthorrefmark{2}, Helmut Neukirchen\IEEEauthorrefmark{3}, Abdulrahman Azab\IEEEauthorrefmark{4}	}
	\IEEEauthorblockA{
	    \IEEEauthorrefmark{1} Natural History Museum, University of Tartu, Estonia, kessy.abarenkov@ut.ee\\
    \IEEEauthorrefmark{2} Department of Geosciences, University of Oslo, Norway, annefou@uio.no\\
	\IEEEauthorrefmark{3} School of Engineering and Natural Sciences,  University of Iceland, helmut@hi.is\\
    \IEEEauthorrefmark{4} 
		Division of Research Computing,
		University Center for Information Technology (USIT),\\
		University of Oslo, Norway,
		azab@uio.no\\
   }
   \thanks{This project has received funding from the European Union's Horizon 2020 research and innovation programme under grant agreement No 857652 EOSC-Nordic.}
}
\maketitle
\copyrightnotice
\begin{abstract}
To reproduce eScience, several challenges need to be solved:
scientific workflows need to be automated; the involved software versions need to be provided in an unambiguous way;
input data needs to be easily accessible; High-Performance Computing (HPC) clusters are often involved
and to achieve bit-to-bit reproducibility, it might be even necessary to execute the code on a particular cluster 
to avoid differences caused by different HPC platforms (and unless this is a scientist's local cluster, 
it needs to be accessed across (administrative) borders). 
Preferably, to allow even inexperienced users to (re-)produce results, all should be user-friendly. 
While some easy-to-use web-based scientific portals support already to access HPC resources, 
this typically only refers to computing and data resources that are local. 
By the example of two community-specific portals in the fields of biodiversity and climate research,
we present a solution for accessing remote HPC (and cloud) compute and data resources from scientific portals across borders, 
involving rigorous container-based packaging of the software version and setup automation, thus enhancing reproducibility.
\end{abstract}
%

\begin{IEEEkeywords}
Reproducibility, Cross-border computing, Workflows, Scientific portals, PlutoF, Galaxy, HPC, Containers
\end{IEEEkeywords}

\section{Introduction}
\label{sec:intro}
The European Open Science Cloud (EOSC)~\cite{ayris2016realising} aims at providing European researchers 
a federated and open multi-disciplinary environment where they can publish, find, and use data, tools, 
and services for research, innovation, and education. The EOSC-Nordic\footnote{\url{https://www.eosc-nordic.eu/}} 
research project aims at fostering EOSC at the Northern European and Baltic level. 
Researchers in different countries and from several scientific disciplines strive to use High Performance Computing (HPC) resources for 
scientific analysis of data.
With such a heterogeneous group of users and HPC resources, reproducibility of scientific workflows is an issue. 
Reproducibility on HPC systems is highly complex with very technical challenges: if different versions of the involved software and dependencies are being used on the same HPC cluster, or if the analysis/simulations are run on different HPC clusters (which is likely to happen considering the lifespan of HPC systems) or smaller machines, the results obtained will be different~\cite{geyer2021limits}. 

The reproducibility of scientific data analysis can be improved by describing all the involved steps and creating automated workflows.
However, most researchers are non-experienced users of complex HPC systems which limits the number of researchers that can reproduce an automated workflow or even create automated HPC workflows.

Web-based scientific portals provide user-friendly interfaces between inexperienced HPC users and HPC systems.
Galaxy~\cite{galaxy} and PlutoF~\cite{Abarenkov2010} are popular scientific portals that can act as a user-friendly web interface between users and HPC systems. 
Such portals provide visual workflow tools to describe and automate scientific workflows involving computational jobs 
that can be converted to HPC job descriptions ready for submission to an HPC cluster where the job processes data. 
By providing the software needed for the computational jobs as virtual environments (virtual machines or containers), 
it can be taken care that the same software versions are used when the workflows are executed by different researchers on different machines, 
thus enhancing reproducibility. 

For bit-to-bit reproducibility of results, it might even be necessary to run on exactly the same HPC cluster~\cite{geyer2021limits} 
which may require access to that HPC cluster by researchers who typically do not have access to it.
However, an
issue when using HPC across (administrative) borders is the management of access control. The portal access to the remote HPC resources is subject to the 
user access and resource quota management on the 
HPC site. 
In addition, cross-border access is required in many cases, i.e.\ portal users in Sweden needing access to an HPC facility in Finland. 
The main obstacle is that it is not trivial to map portal users to HPC users due to technical security and data protection barriers. 
One option is to run portal jobs as one single anonymous user 
for \emph{all} users of 
this particular portal, but this would introduce quota management issues for the different 
portal
user groups. 

This paper presents a solution 
to support submission of jobs from community-specific portals, taking Galaxy and PlutoF as examples, to a variety of HPC systems. 
The solution lies in the creation of a \textit{robot user} at each HPC facility for \emph{each} group of users, so that user portals submit jobs as this robot user. 
On the HPC side, each robot user is associated with a user group to which a specific quota is assigned and managed. 
Also, the data to be processed via a portal on an HPC cluster might be stored elsewhere, e.g., 
in public data repositories, and the portal may not support 
access to the data. Therefore, we had to enhance the portals to allow accessing remote data.
This enhances the FAIRness (Findable, Accessible, Interoperable and Reusable) of the researchers' work~\cite{Wilkinson2016}. 

Our solution has been evaluated using two EOSC-Nordic pilot case studies, \textit{Biodiversity} and \textit{Climate}, 
which are described in sections~\ref{sec:biodiversity} and~\ref{sec:climate}, respectively.
Automated workflows (including pre- and post-processing) for HPC (or even cloud-based systems, e.g., for High Throughput Computing) 
on distributed data and computing resources across borders enable community-specific or thematic portals 
(which are traditionally designed to submit jobs only on local clusters) to execute jobs in a reproducible manner even on remote resources by packaging software using virtualisation technology. 
While we describe our approach using two case studies, the approach itself is generic, i.e.\ independent from the architecture and the technology of any given portal and the target HPC queuing system.
A summary and outlook are provided in Section~\ref{sec:conclusions}.


\section{Biodiversity Pilot}
\label{sec:biodiversity}

The Biodiversity pilot supports researchers from fields such as molecular ecology, taxonomy, or biodiversity with 
species~discovery from 
environmental DNA (eDNA) samples, and unambiguous and traceable communication of these taxa. 

Analysis tools for a large amount of molecular sequence data can require a significant amount of HPC resources. 
Setting these tools up in an HPC cluster is not easy as it requires knowledge in bioinformatics and information technology.
Instead, our PlutoF~\cite{Abarenkov2010} portal makes digital services of UNITE~\cite{Nilsson2018}, 
a database and sequence management environment,
available for the UNITE user community by providing a simple front-end solution as an alternative to a command line interface. 
PlutoF is a web-based workbench\footnote{\url{https://plutof.ut.ee}} and computing service provider for biology and related disciplines. 
It features an analysis module by providing services for molecular sequence identification and species discovery from eDNA samples.
PlutoF handles user management, logging, and storing analysis runs and data files, while executing jobs on the local
University of Tartu UT Rocket HPC cluster~\cite{https://doi.org/10.23673/ph6n-0144}.

To be able to provide more HPC resources based on the individual user's needs and improve the reproducibility of researchers' workflows, we enhanced our PlutoF platform to integrate any other non-local HPC resource where users have access to, 
thus allowing analysis jobs to be submitted on these remote (and typically cross-border) resources. 

The main goals in the Biodiversity pilot were: 
\begin{enumerate}
    \item 
  Package services in a way that allows service providers to easily build, transfer, and run these services independent of the software available in remote HPC clusters, thus enabling reproducibility;
   \item 
  Allow service providers to send PlutoF analysis jobs to remote EOSC-Nordic HPC clusters (given that there is a user community with access to 
 this cluster);
    \item 
  Work out a recommended procedure on how users can apply for and access EOSC HPC resources from PlutoF in a standard, consistent, simple, and automated way.
\end{enumerate}

\subsection{Resources}
\label{sec:biodiversity-resources}
Based on the PlutoF user community, we identified two HPC providers to test our cross-border computing implementation: 
\begin{inparaenum}[a)]
\item the Swedish National Infrastructure for Computing (SNIC) 
and 
\item the National Institute of Chemical Physics and Biophysics that provide the NICPB HPC cluster 
as part of the Estonian Scientific Computing Infrastructure (ETAIS).
\end{inparaenum}

For testing, the Swedish user community did set up small projects at the SNIC Science Cloud (SSC) and the SNIC High Performance Computing Center North (HPC2N), which are connected from the PlutoF resource via ssh. The ssh account was used to test popular PlutoF analysis services for a massive data set from the Artificial Reef Monitoring Systems (ARMS) in the Baltic and the North Sea~\cite{Obst2020}. 

We as PlutoF team had a list of prerequisites for the resource providers: 
\begin{enumerate}
\item SLURM~\cite{slurm} workload manager, 
\item Singularity (recently renamed to: Apptainer) container, 
\item robot account for submitting jobs allowed. 
\end{enumerate}

Not all candidate HPC providers were able to meet these requirements. For example, the SNIC User policy does not allow submitting jobs as robot user to SNIC's HPC2N cluster, 
although such a case is allowed in SNIC's SSC cloud. The SSC, on the other hand, comes as a Virtual Machine (VM) without any software installed and thus, 
an automated setup of a Virtual Research Environment (VRE) to install required software was needed (with the benefit of enhanced reproducibility).

\subsection{Technical Solutions}
\label{sec:biodiversity-tech}

\subsubsection{Front-end}
\label{sec:biodiversity-front-end}

As part of the pilot, functionality for sending jobs to different remote HPC clusters 
(instead of the local UT Rocket cluster only) was added to PlutoF.
This required support for switching HPC resource providers at the user level 
(based on the user's preference and availability of HPC providers), 
and setting proper access parameters when submitting jobs to and receiving job results from a remote resource. 

Analysis data files and SLURM scripts are copied (via ssh, scp, or rsync through ssh tunnels) 
from PlutoF to the remote HPC cluster, jobs are started and executed remotely, 
and analysis results are fetched by PlutoF once jobs are finished. 
Users are notified upon job completion via email.

\subsubsection{Packaging}
\label{sec:biodiversity-packaging}

To enhance reproducibility, capability for automated building and installation of the needed software was added by packaging 
PlutoF digital services into Singularity containers with container building code, 
automated setup scripts, and documentation available in GitHub.

\subsubsection{Setup automation}
\label{sec:biodiversity-setup}

The process of wrapping PlutoF digital services into Singularity containers was documented 
and published as GitHub repositories, and can be used to automate the installation process, thus enabling reproducibility. 
The automated setup scripts
cut down the installation time from several hours to approximately 10 minutes in total. 
This includes four digital services to support the eDNA-based species discovery:
\begin{inparaenum}[a)]
   \item ITSx\footnote{\url{https://github.com/TU-NHM/itsx_plutof_pub}} (detection and extraction of ITS1 and ITS2 from ribosomal internal transcribed spacer (ITS) sequences),
    \item PROTAX-fungi\footnote{\url{https://github.com/TU-NHM/protax_fungi_plutof_pub}} (taxonomic placement of fungal ITS sequences),
    \item massBLASTer\footnote{\url{https://github.com/TU-NHM/massblaster_plutof_pub}}, and
    \item SH matching analysis\footnote{\url{https://github.com/TU-NHM/sh_matching_pub}}
.
\end{inparaenum}

\subsubsection{Potential blockers and sustainability issues}
\label{sec:biodiversity-sustain}

We identified the following issues during our work:
\begin{enumerate}

    \item ssh key-based authentication was not supported by all resource providers, e.g.\ SNIC HPC2N supports only Kerberos/GSSAPI-based authentication which we needed therefore to implement in PlutoF.
    \item HPC service access is normally provided for a certain time period, after which the user has to go through the process of applying for resources again.
    \item Robot user accounts are often not allowed. This was the case with SNIC HPC2N, so would could only support single-user accounts.
    \item Constant maintenance (e.g. software and operating system updates, Singularity container updates, 
    resolving VM service interruptions, and unexpected failures) of the VMs (for SSC and similar cloud-based cases) 
    requires additional work and resources from the technical team.
    \item Access to Nordic HPC resources for \emph{all} PlutoF platform users is impossible to implement: access in EOSC-Nordic HPC clusters requires belonging to an HPC project which has been given access with limited resource quota -- this is currently not the case for all PlutoF users.
\end{enumerate}

\subsection{Data-flow and Benchmarking}
Input data is uploaded for each workload submission while big reference data is shipped once together with the Singularity container that includes the actual toolbox.
As an example, data transfer overhead for an actual compute job (processing SH matching analysis) on the PlutoF platform using UT Rocket HPC cluster is presented in Table~\ref{tab:table1}.

\begin{table}[t!]
  \begin{center}
    \caption{Benchmarking results for the SH matching analysis via the PlutoF platform. Pre-processing (column Pre-proc) includes input data transfer from PlutoF server to HPC and was just a few seconds (rounded to 0 min). Processing (Proc) includes time for running the analysis. Post-processing (Post-proc) includes the transfer of analysis results back to PlutoF server, updating job status, and sending out email notification to the user. Post-proc is largely dependent on a crontab process where the presence of analysis results in the HPC cluster is checked periodically every 10 minutes.}
    \label{tab:table1}
    \begin{tabular}{rSccc}
    \toprule
      \textbf{Records} & \textbf{File size} & \textbf{Pre-proc} & \textbf{Proc} & \textbf{Post-proc}\\
      count & kB & min & min  & min \\
      \midrule
      10 & 5.542 & 0 & 12 & 0\\
      100 & 58.108 & 0 & 118 & 2\\
      1\,000 & 585.87 & 0 & 148 & 9\\
      10\,000 & 5827.32 & 0 & 205 & 9\\
      100\,000 & 16380.5 & 0 & 552 & 6\\
    \bottomrule
    \end{tabular}
    \vspace*{-2ex}
  \end{center}
\end{table}

\subsection{Take-up}
\label{sec:biodiversity-take-up}
Since May, 2020 when UNITE services were moved to Singularity containers, 2120 analysis runs by 260 users  (data from January 17, 2022) have been started in PlutoF.
In April 2021, a PhD course linked to an open workshop about building the forest biodiversity open data services was organised by the NEFOM network. 
UNITE digital services were presented at that workshop and taught during that PhD course.

The use of PlutoF and the described improvements enable researchers to easily develop automated workflows and make their scientific data analysis Open and Reproducible. 

\section{Climate Pilot}
\label{sec:climate}

\begin{figure*}[t]
\centering
\includegraphics[width=0.9\textwidth,keepaspectratio=true]{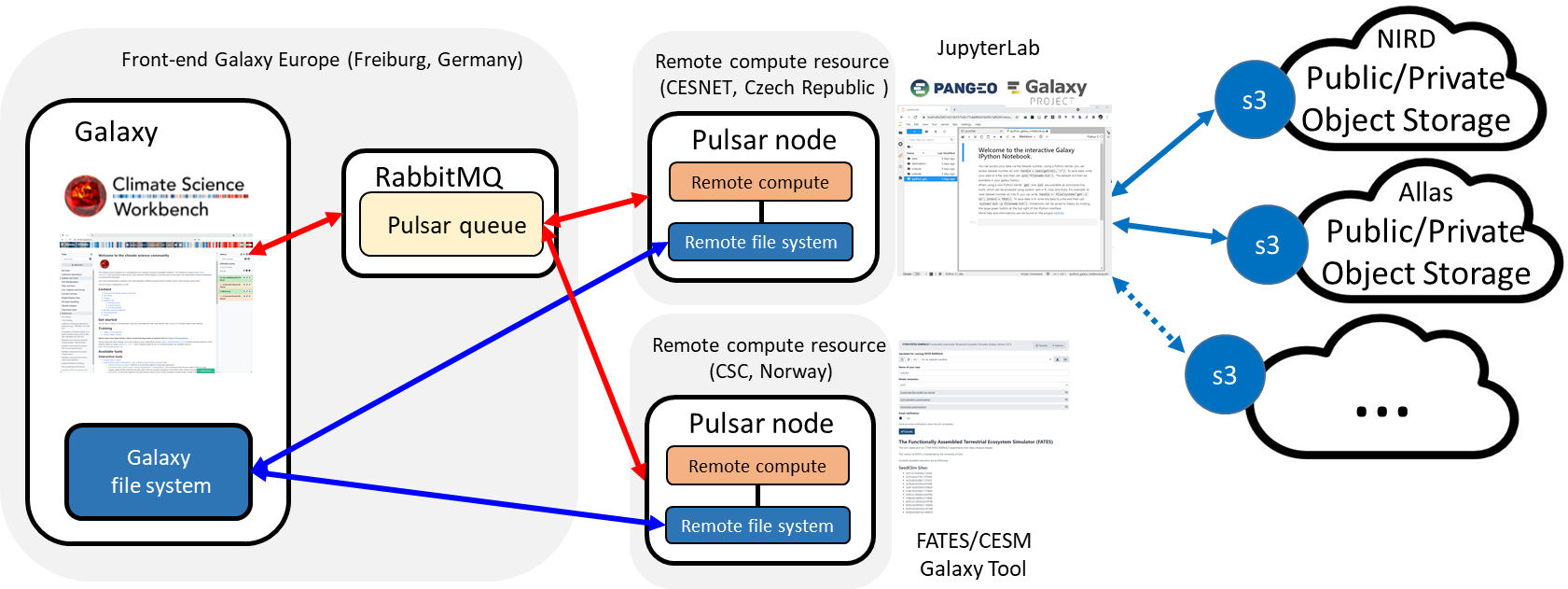}
\caption{Overview of the different components of Galaxy Climate: i) left-hand side: Galaxy Climate front-end; ii) center: Remote compute resources are added (new Pulsar node) and are selected depending on their availability and tool requirements, e.g.\ GPUs, memory, etc. iii) right-hand side: Object Storage end-points to access data remotely and independently of their physical locations.}\vspace*{-2ex}
\label{fig:demo}
\end{figure*}

Research related to climate change is intrinsically interdisciplinary and entails significant scientific and technical challenges \cite{Schipper2021}. 
One example is the development and use of Earth System Models (ESMs): to improve the transparency and reproducibility of climate experiments, the same source code and the same processor layout as well as the same computational environment (compilers including optimization flags, libraries such as MPI or netCDF) are needed. Facilitating the development of fully automated workflows for running ESMs is key to enable scientists to create fully reproducible simulations and/or to easily reproduce simulations. 

The Climate pilot is based on the ecosystem of the Galaxy portal~\cite{galaxy} and to achieve the above goals, we had to: 
\begin{enumerate}
 \item Package climate tools following the EOSC-Life methodology framework to enhance reproducibility \cite{Bietrix2021} 
   and develop the corresponding Galaxy tools to offer a Graphical User Interface (GUI) to end-users for developing and running fully automated workflows;
  \item Allow climate simulation and analysis jobs to be sent to remote and cross-border EOSC-Nordic compute resources, 
  including HPC resources (for the latter, a user community with access to resources in the respective HPC cluster is necessary) 
  and maintain bit-to-bit reproducibility (so that ESM outputs obtained on various machines are identical for a given domain decomposition);
 \item Provide remote access to storage resources (S3-compatible object storage) to share input/output data to limit copies and run efficiently ESM workflow tasks on different computing resources, 
  as independently as possible from the storage location;
 \item Work out recommended procedures on how EOSC HPC resources can be added in Galaxy.
\end{enumerate}

\subsection{Technical Solutions}
\label{sec:climate-tech}

Galaxy~\cite{galaxy} is an open-source platform for FAIR data analysis enabling scientists to develop fully automated and reproducible workflows to analyse data with minimal technical impediments. 
It supports pluggable inter-operable tools, graphical workflow editing, visualisations, integrated 
training infrastructure, and has an active community. Free online analysis is supported, 
running on large scale US, European, and Australian research computing infrastructures, available\footnote{\url{https://github.com/galaxyproject/galaxy}} 
as source code or as container images for desktop, local cluster, or cloud deployment. The Galaxy Climate portal used in this pilot is deployed on Galaxy Europe\footnote{\url{https://usegalaxy.eu/}} but any automated workflow could run on any other Galaxy portal while providing fully reproducible results.

\subsubsection{Front-end}
\label{sec:climate-front-end}

Our overall setup of Galaxy Climate is shown in Fig.~\ref{fig:demo}:
Pulsar is the Galaxy Project’s remote job execution system 
that allows Galaxy instances to execute jobs remotely (there is no need to have a shared file system). 
A Galaxy instance sends all the data necessary to execute a job to Pulsar which handles the part of installing and preparing all the tools (also called 
``staging''), scheduling the jobs, etc. After the computations have been completed, the results are sent back to Galaxy Climate.
On the back-end side, Pulsar (like Galaxy) can use various job-managers 
(any Distributed Resource Management Application API (DRMAA)-compliant~\cite{Troger2007}, but also, e.g.,\ SLURM) depending on the target machines (cloud computing, HPC cloud, or bare metal HPC).

Various object storage end-points where data can be accessed either privately or publicly have been added to allow transparent data access from any computing resource. 
S3-compatible object storage end-points are accessible via any Galaxy tool, including, e.g., interactive Jupyter Notebooks. 
In addition, when uploading a large dataset in the Galaxy user space, end-users can choose to defer the dataset resolution: 
in that case, the dataset is directly uploaded on the target machine where the tool effectively runs, thus optimizing data transfer. 

\subsubsection{Packaging}
\label{sec:climate-packaging}
The first step to getting a new climate tool deployed into Galaxy Climate is to develop a 
conda (a cross-platform package and environment manager)\footnote{\url{https://conda.io}} package for it. 

The second step is to create the Galaxy wrapper that describes all inputs, outputs, and parameters of a tool, 
so that Galaxy generates a GUI out of it and subsequently a command to be sent to the cluster. 
A Galaxy wrapper is an XML file containing the description of the requirements 
(conda packages and versions for the tool itself and all the dependencies needed for the execution of the tool), inputs and outputs, 
and most importantly annotations to make Galaxy tools FAIR. 
All the Galaxy Climate tool wrappers are published in the Galaxy ToolShed\footnote{\url{https://toolshed.g2.bx.psu.edu/}} under the ``Climate Analysis'' 
category and maintained by the Nordic ESM Hub in the repository \emph{galaxy-tools} on Github\footnote{\url{https://github.com/NordicESMhub/galaxy-tools}}. 

Finally, a bot automatically creates (Bio)Containers\footnote{\url{https://github.com/BioContainers/specs}} 
(Docker, RKT, and Singularity) by tracking all Galaxy tools to ensure that a container exists for each tool. 

\subsubsection{Setup automation}
\label{sec:climate-setup}

Once a new climate tool is available in the Galaxy ToolShed (and the corresponding containers have been automatically created), 
it can be installed on any Galaxy server. On Galaxy Europe, a Pull Request to the \emph{usegalaxy-eu-tools} Github 
repository\footnote{\url{https://github.com/usegalaxy-eu/usegalaxy-eu-tools}} is required for installing new tools 
while tool upgrades for the Climate Community are done automatically: a new version of a given tool is installed 
whenever it becomes available in the ToolShed, but the Galaxy server keeps all the previous versions to make sure existing workflows are fully reproducible.

\subsubsection{Potential blockers and sustainability issues}
\label{sec:climate-sustain}

We identified the following issues during our work:
\begin{enumerate}
    \item Dedicated, targeted training (with high-quality online training materials) is paramount to on-board new users or show new functionalities to existing, experienced users. This will also increase the number of fully reproducible automated workflows published for instance in the WorkflowHub registry\footnote{\url{https://workflowhub.eu}}, a registry for describing, sharing and publishing scientific computational workflows.
    \item Exchanging data and histories with colleagues using different Galaxy instances is cumbersome, especially when data is large (which is typically the case for Climate analysis) and can hinder practical reproducibility (data too large and slow to access to be easily reproduced). For this reason, users are encouraged to migrate the results of their simulation to an object storage that can be publicly accessed outside the Galaxy portal. Eventually, the results may be moved to an archive, so it is important to add extensive metadata, persistent identifiers, and to create data catalogs (such as zarr catalogs) for improving the usage and the reproducibility of the results. The usage of Object Storage is already available in Galaxy (users can easily save the data they upload or produced in an available object storage) and tools to migrate to an archive and create the associated data catalog will be developed.
    \item Using cloud-optimized formats (e.g.\ zarr) is often necessary but trade-off between performance and ease of access (for instance from a laptop) is necessary and non-trivial. 
    \item While ESMs in containers are fully reproducible (bit-for-bit reproducibility with the same configuration, typically even on different HPC and/or cloud resources) and perform extremely well on bare-metal HPC, adding bare-metal HPC resources to a Galaxy instance (as a new Pulsar node) 
    raises security and performance issues and requires further development of the Pulsar service. 
    The Horizon Europe project EuroScienceGateway (starting in September 2022) will address these issues and
    lift Pulsar from technology readiness level \mbox{TRL-7} to \mbox{TRL-9} by expanding the APIs, 
    hardening deployments, and adding support for different usage patterns, e.g.,\ data localisation during job scheduling.
\end{enumerate}

\subsection{Take-up}
 Training material on climate science for on-boarding users has been developed and is publicly available online and maintained by the Galaxy Training
 Network\footnote{\url{https://training.galaxyproject.org/training-material/topics/climate/}}. Online training events are regularly organized to teach researchers how to develop fully automated and reproducible analyses with Galaxy. Galaxy Climate
 Community meetings are regularly organized to present the recent updates (new Galaxy tools and functionalities) and take inputs from the community. 
 To widen the usage of the Galaxy Climate data analysis
 workbench beyond the Nordic and Baltic countries, the Galaxy Climate community is taking part in Outreachy\footnote{\url{https://www.outreachy.org}} that provides internships in open source and open science
 to people subject to systemic bias and impacted by under-representation in the technical industry where they are living. This allows to promote open science and reproducible research at a larger scale.

\section{Conclusions}\label{sec:conclusions}
We presented a user-friendly solution enabling scientists to use cross-border cloud and HPC resources through web portals while achieving reproducibility using containers and workflow automation. 
The solution has been developed by the EOSC-Nordic project and was applied and tested using two pilots. We successfully managed to: 
\begin{inparaenum}[a)]
  \item Package data analysis services and tools according to best practices, so that they can be deployed and executed in a reproducible manner on different compute facilities (independent of from the installed software and, e.g.,\ the HPC queuing system); 
  \item Develop technical solutions and recommended procedures on how their services can be coupled with cross-border compute resources and remote data.
\end{inparaenum}

The selected pilots have explored different technical solutions to serve their respective community and the take-up of the services by researchers has been very good. 
However, both faced an administrative issue related to the use of robot accounts with bare metal HPC and while technical solutions have been proposed, 
their implementation would require changes that need to be handled at an administrative policy level. 

Another issue faced by both pilots, was the sustainability of the services made available at remote HPC clusters. 
HPC service access is normally provided for a certain time period, after which the user has to go through the process of applying for resources again. 
This constant renewing of access and change of the HPC providers' specifics requires a number of actions at both sides -- at the user and at the service provider. 

Also, to support job managers, such as Flux~\cite{Flux0202}, that are tailored to the heterogeneous resources of modern exascale systems, more detailed job resource descriptions would need to be added to the portals. 

\subsection{Next steps for the Climate pilot}

The following improvements are planned:
\begin{inparaenum}[a)]
    \item Plug-in bare-metal HPC resources to Galaxy (including the European pre-exascale HPC system LUMI\footnote{\url{https://www.lumi-supercomputer.eu/}}, where actual tests show containerized ESMs are scaling with performance similar to bare-metal runs and still bit-for-bit reproducibility) to run higher resolution and longer simulations; this will be implemented with our robot user approach.
    \item Reduce data movement by improving Pulsar or directing certain climate jobs to specific Pulsar nodes where the corresponding climate data is available.
\end{inparaenum}

\subsection{Next steps for the Biodiversity pilot}
We 
will continue testing the two different remote platforms (SNIC SSC and SNIC HPC2N) for improving the tools and adjusting environment parameters according to the users' needs. 
As a next step, we will use online media channels and scientific articles to advertise the cross-border computing solution and the possibility of linking new HPC resources to PlutoF.
We are currently organizing PhD courses to present and teach the tools made available through PlutoF as EOSC-Nordic services. 
Although we successfully coupled the PlutoF platform with the SNIC HPC cluster using a robot account, 
we have only implemented a test solution where we apply a robot account for one single user. 
This process needs to be elaborated further to become a sustainable solution accepted by SNIC.
As part of this, we are also planning to add UNITE digital services provided by PlutoF as an EOSC service.

\bibliographystyle{IEEEtran}
\IEEEtriggeratref{7} 
\bibliography{stroll}
\end{document}